\documentclass{elsart}
\usepackage{natbib}
\usepackage{epsfig} 
\begin{document}
\runauthor{Wang}
\begin{frontmatter}
\title{Beam Test of BTeV Pixel Detectors}

\author[Syracuse]{J.C. Wang\thanksref{correspond}},
\author[Fermi]{J.A. Appel},
\author[Syracuse]{M. Artuso},
\author[Fermi]{J.N. Butler}, 
\author[Fermi]{G. Cancelo},
\author[Fermi]{G. Cardoso}, 
\author[Fermi]{H. Cheung}, 
\author[Fermi]{G. Chiodini},
\author[Fermi]{D.C. Christian},
\author[INFN]{A. Colautti},
\author[Milan]{R. Coluccia},
\author[Milan]{M. Di Corato},
\author[Fermi]{E.E. Gottschalk}, 
\author[Fermi]{B.K. Hall}, 
\author[Fermi]{J. Hoff}, 
\author[Fermi]{P. A. Kasper},
\author[Fermi]{R. Kutschke}, 
\author[Fermi]{S.W. Kwan}, 
\author[Fermi]{A. Mekkaoui},
\author[INFN]{D. Menasce}, 
\author[Iowa]{C. Newsom}, 
\author[INFN]{S. Sala},
\author[Fermi]{R. Yarema}, and
\author[Fermi]{S. Zimmermann}

\thanks[correspond]{Corresponding author}

\address[Fermi]{Fermi National Accelerator Laboratory, Batavia, IL 60510, USA}
\address[Iowa]{University of Iowa, Iowa City, IA 52242, USA}
\address[INFN]{Sezione INFN di Milano, via Celoria 16-20133 Milano, Italy}
\address[Milan]{Universit\'{a} di Milano, Dipartimento di Fisica, 
via Celoria 16-20133 Milano, Italy}
\address[Syracuse]{Syracuse University, Syracuse, NY 1344-1130, USA}

\begin{abstract}
The silicon pixel vertex detector is one of the key elements of the BTeV spectrometer.
Detector prototypes were tested in a beam at Fermilab.
We report here on the measured spatial resolution as a function of the incident angles
for different sensor-readout electronics combinations.
We compare the results with predictions from our Monte Carlo simulation.
\end{abstract}


\end{frontmatter}

\section{Introduction}
BTeV is an experiment being constructed at the Fermilab $p\bar{p}$ collider,
designed to study mixing, CP violation, and rare decays of beauty and charm hadrons.
This experiment exploits two advantages of the ``forward'' direction: the correlation in
the direction of the $b\bar{b}$ pair produced, and the boost that allows an easier
identification of detached vertices.
This allows for efficient flavor tagging and sensitivity to a great variety of heavy
flavor decays~\cite{btev}. 

The pixel vertex detector is one of the key elements of the BTeV spectrometer.
It is located in the center of the detector, inside an 1.6 T dipole magnet.
The pixel technology is chosen for its higher radiation resistance and superior pattern
recognition capability, which is essential to the detached vertex trigger~\cite{erik},
that depends extremely on pixel information on the first level.

During 1999-2000 fixed target run at Fermilab, several single chip pixel detector 
prototypes were tested, with different front end readout chips.
The goal was to determine the spatial resolution as function of particle angle of incidence,
readout digitization accuracy, sensor bias and readout thresholds.
We also wanted to gain operational experience and to look for potential problems.

A detailed Monte Carlo (MC) simulation was developed to predict spatial resolution and occupancy
for different sensors and front-end chips.
MC and data are compared below to test our understanding of these detectors.

\section{The experimental apparatus}
\begin{figure}[htb] 
\centerline{\epsfig{figure=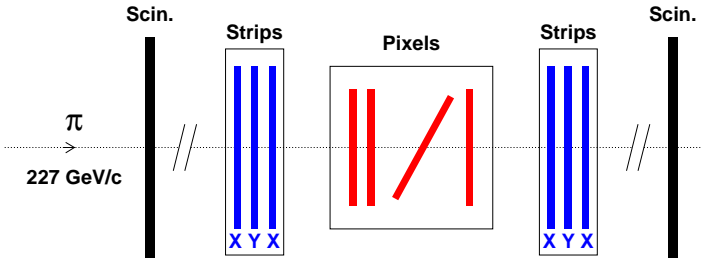,height=1.4in}} 
\caption{ \label{telescope}Schematic drawing of the telescope.} 
\end{figure}

The data were collected using a 227 GeV/c pion beam.
Figure~\ref{telescope} shows the experimental setup.
The pixel devices under test were located between two stations of silicon
microstrip detectors (SSD's),
that provide tracking information, to an accuracy of about 2.1 $\mu$m in the x direction
on pixel planes, that corresponds to the ``small pixel dimension''.
The pixel cell is rectangular ($\rm 50 \mu m \times 400 \mu m$).
In the resolution we report, the contribution from tracking precision is not taken out.
The pixel devices were mounted on printed circuit boards (PCB's),
which fit tightly into slots machined in an aluminum box.
The slots allowed a single detector to be positioned normal to the incident beam direction or
at different angles.

The pixel sensors under test were all from the ``first ATLAS prototype submission''~\cite{atlas}.
The pixel collection electrodes are $\rm n^+$ doped on n substrate,
with p-stop or p-spray insulation techniques.
Different type of sensors were indium bump bonded to two different versions of the front
end readout chip developed at Fermilab~\cite{dave2}.
Two sensors of p-stop and p-spray types from CiS were bump bonded to FPIX0 chip,
and two from Seiko were bump bonded to FPIX1 chip.
Both chips provide zero-suppressed analog information.
The FPIX0 output was analog and digitized by an external 8 bit flash ADC (FADC).
The FPIX1 included a 2-bit FADC in each cell.

Two scintillation counters were positioned upstream and downstream of the silicon telescope.
The readout was triggered by the coincidence of signals from these two scintillation counters,
and the $\rm FAST\_OR$ output signal from the FPIX0 p-stop detector.

\section{Monte Carlo Simulation}
A detailed Monte Carlo simulation was developed to predict the charge sharing
and spatial resolution of silicon pixel detector~\cite{marina}.
We compared the simulation with beam test results to determine the validity of the
simulation model.

In an $n^+np^+$ detector, the charge carriers collected at the pixel electrodes are 
electrons.
A minimum ionizing particle (MIP) crossing silicon sensor, deposits energy
along its path, producing electron-hole pairs.
The charge clouds drift along electric field, and spread laterally due to diffusion.
The mobility dependence upon electric field and temperature is included in the
simulation~\cite{cms}.
In simulating the energy deposition, we take into account the atomic structure of the
material~\cite{geant}.
The processes determining the energy loss are: excitation, low energy ionization and
energetic knock-on electrons ($\delta$-rays).
The treatment of $\delta$-rays is similar to reference~\cite{pindo}.

Pixel detectors feature very low electronic noise.
In particular, the equivalent noise charge (ENC) is about 110 $e^-$ for FPIX0-pstop
and FPIX1-pstop.
The external FADC introduces extra noise of about 400 $e^-$~\cite{gabriele}.
The discriminator thresholds are 2500 and 3780 $e^-$ for FPIX0-pstop and FPIX1-pstop
respectively, with spread of about 400 $e^-$.
The effects of amplifier, threshold and ADC accuracy are included in the simulation.

\section{Results}
\subsection{Charge collection}
Charge collection is studied with FPIX0-instrumented sensors, because of the 8-bit
digitization resolution.
The measured pulse height distributions are well fitted using a Landau distribution convoluted
with a Gaussian.
For the CiS p-spray sensor, we find sizable charge collection inefficiency on the column
boundaries.
This is consistent with previous measurements made by the ATLAS collaboration~\cite{ragusa}.
This feature is attributed to a design flaw and is not intrinsic of the p-stop technique.

The average charge collected by CiS p-stop sensor is about 30,100 $e^-$,
with a most probable charge of 24,700 $e^-$.
For CiS p-spray, we observe a 24\% inefficiency in charge collection.
The average charge is about 23,100 $e^-$ and most probable charge is 20,000 $e^-$.

\subsection{Charge sharing}
The charge deposited by a single track is often shared by more than one pixel.
The amount of charge sharing is determined by the number of pixel cells crossed by
the  track, and by the diffusion induced spread of the electrons drifting in the silicon.

\begin{figure}[htb] 
\centerline{\epsfig{figure=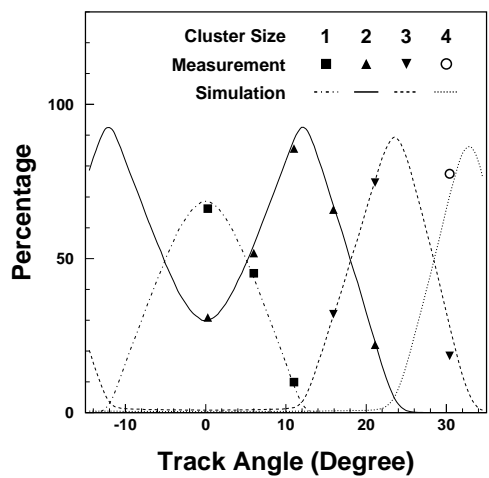,height=3in} }
\vspace{-.6cm} 
\caption{ \label{charge_share}Measured charge sharing of CiS p-stop sensor bump bonded to
FPIX0 readout chip shown as symbol. The curves are results of our simulation.}
\end{figure}

The fraction of the time that each cluster size (number of pixels in narrow dimension)
is observed is shown in Figure~\ref{charge_share} for FPIX0-pstop.
The data shows clearly the dominance of two pixel clusters within the nominal BTeV
acceptance of 18 degrees.
The MC simulation gives an excellent description of the data.

We now study the influence on the charge sharing of different detector bias
voltages and readout threshold settings.
For large incident angle tracks, charge sharing is mainly determined by geometry;
the effect of bias voltage is negligible.
At small angles, the charge sharing is affected by diffusion. 
Low bias voltage results in longer collection time, and thus increases the charge sharing.
As will be discussed later, this improves the spatial resolution at small angles.
As expected, the discriminator threshold is important at all track angles.
Lower threshold always translates directly into more precise position interpolation.

\subsection{Spatial resolution}
The track parameters are fitted using SSD's and other pixel detectors in the telescope.
The projection point of the track on the pixel plane under study is used as
the predicted position.
The measured coordinate is given as the center of gravity of the pixel
cluster associated with the track.
For clusters with more than 1 pixel, a linear $\eta$ correction is applied.
For clusters with 2 pixels,  $\eta$ is defined as the relative charge difference of the
2 pixels. For clusters with more pixels, charges of head and tail pixels are used.

\begin{figure}[htb] 
\centerline{\epsfig{figure=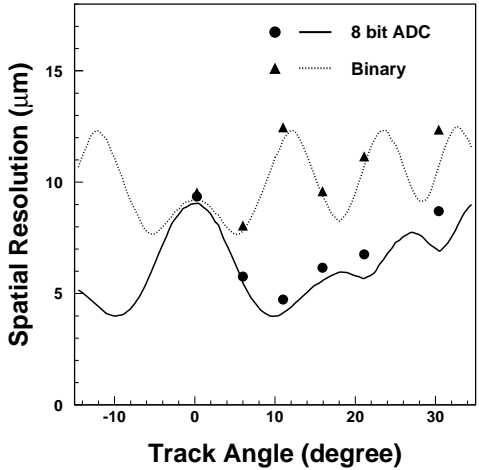,height=3in}}
\vspace{-.6cm}
\caption{ \label{res_fpix0_pstop}Spatial resolution of p-stop sensor bump bonded to
FPIX0 readout chip, shown in symbol are measurement, and curve are simulation. }
\end{figure}

For different track incident angles, the difference between predicted and
reconstructed position (residual) distributions are studied.
The measured spatial resolutions are derived as standard deviations obtained from a Gaussian
fit to these distributions.
The resolutions as function of incident angle are shown in Figure~\ref{res_fpix0_pstop}. 
The experimental results are in very good agreement with the simulation.
The small discrepancy comes from imperfect alignment.

We have computed the resolution with binary information from each pixel hit, also shown in
Figure~\ref{res_fpix0_pstop}.
One interesting feature in binary mode is the oscillation of spatial resolution as the 
incident angle changes.
This oscillation comes from the change of dominant cluster size shown in 
Figure~\ref{charge_share}.
This is also the reason why for small incident angle tracks, the resolution improves
with the decreasing of the sensor bias voltage above depletion voltage with analog readout.
These results show why an analog readout is the chosen solution for BTeV.

\begin{figure}[htb] 
\centerline{\epsfig{figure=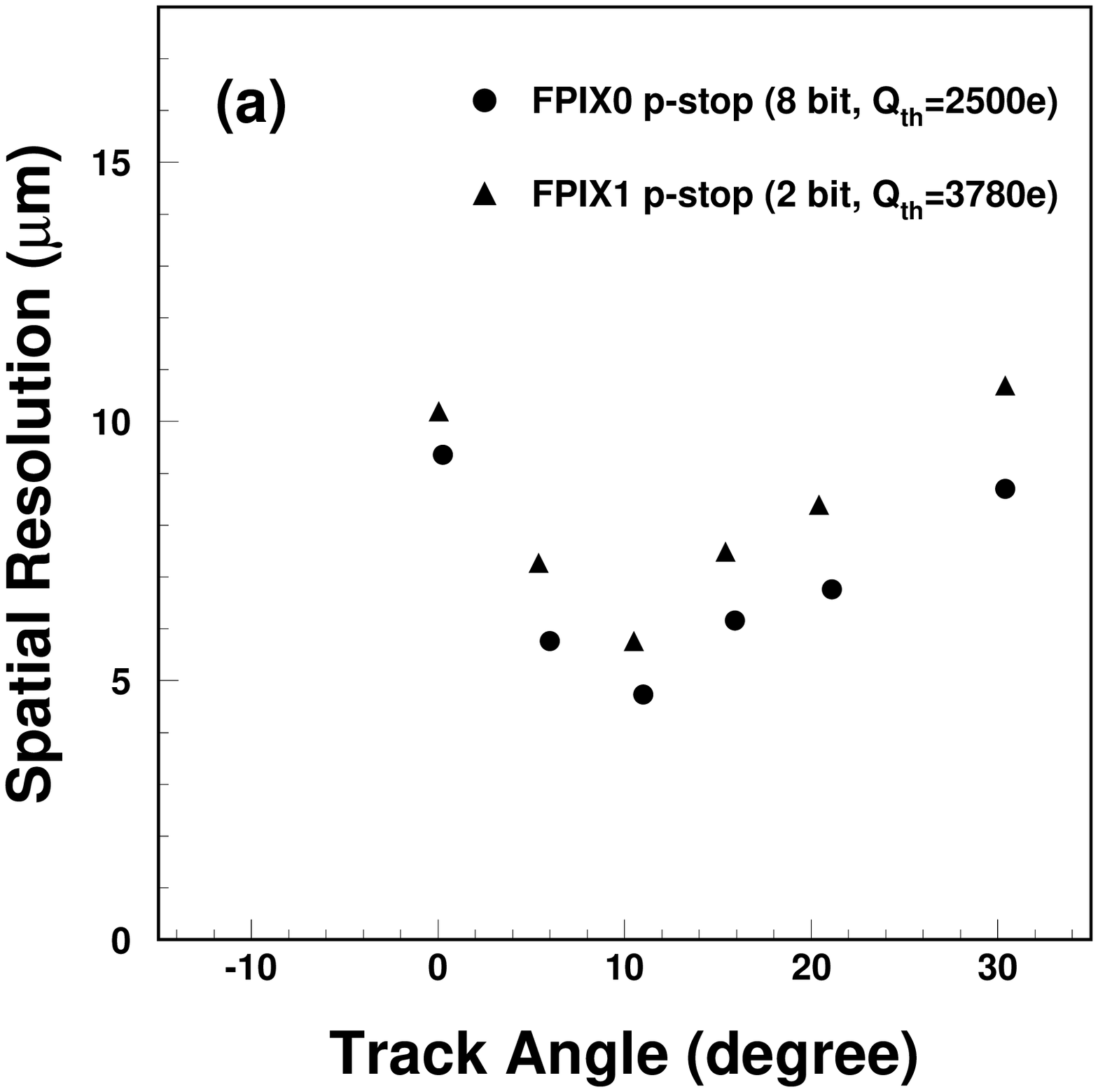,height=3in}
            \epsfig{figure=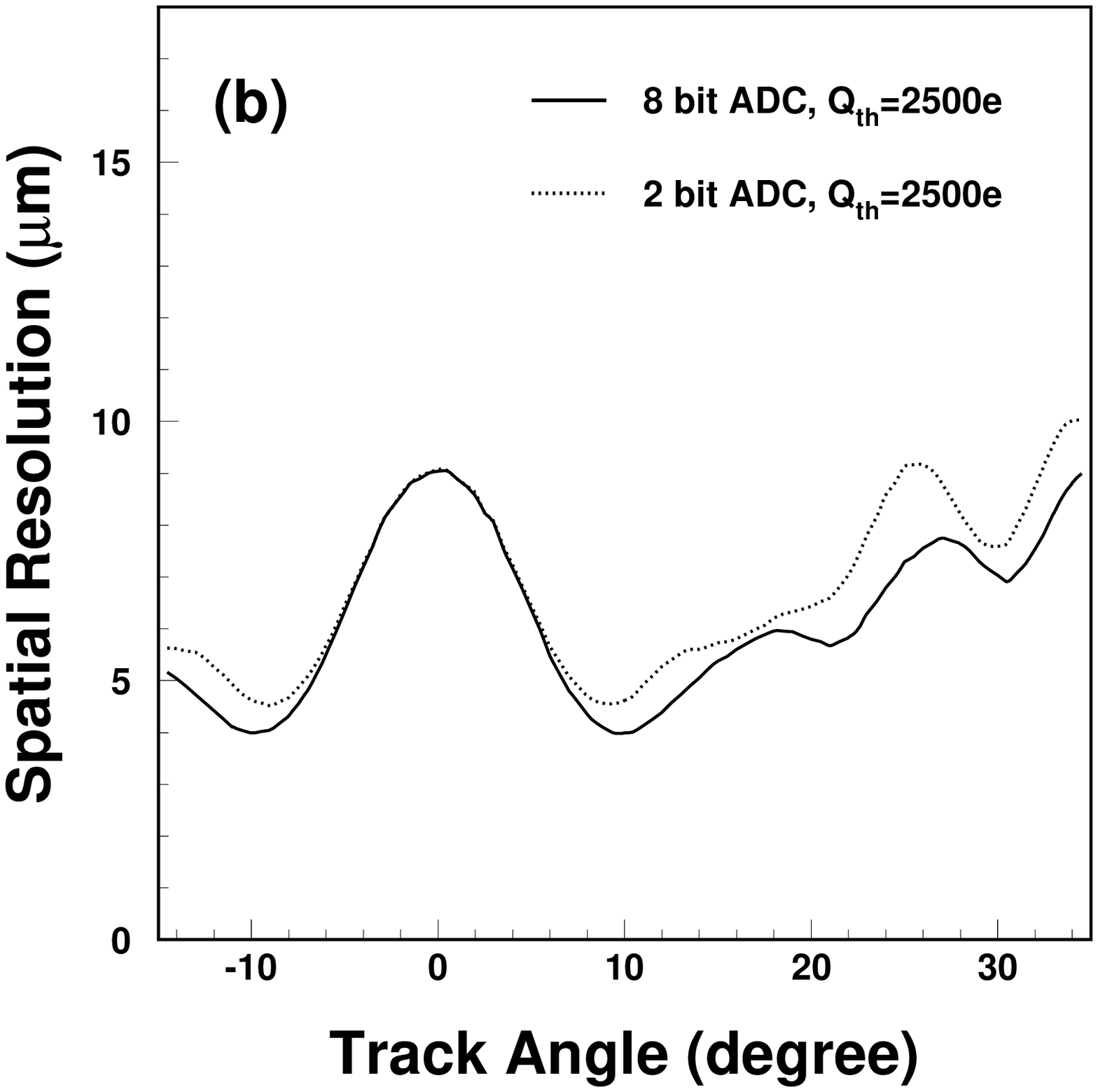,height=3in} }
\vspace{-.6cm}
\caption{ \label{res_all_pstop} Comparisons of spatial resolution with different digitization accuracy:
(a) Measurements with FPIX1 2-bit instrumented p-stop detector (triangle),
and FPIX0 instrumented detector (solid dot). The discriminator thresholds are different. 
(b) MC simulated resolution for 8-bit (solid line), and 2-bit (dotted line) with same threshold.  }
\end{figure}

The spatial resolutions of FPIX1 p-stop sensor are shown in Figure~\ref{res_all_pstop}(a),
The resolution of FPIX1 p-stop is about 1 $\mu$m worse than FPIX0 p-stop.
Most of the deterioration comes from the higher threshold.
The FPIX1 instrumented detector was operated with a discriminator threshold of 3780  $\rm e^-$.
while the FPIX0-instrumented detector was operated with a discriminator threshold of
2500 $\rm e^-$.
Figure~\ref{res_all_pstop}(b) shows the MC simulation results with the same threshold
for 8-bit and 2-bit ADC.
The 2-bit ADC results are quite good and give us confidence that the 3-bit ADC planned for our
final readout electronic (FPIX2)~\cite{abder} will provide excellent performance.

\subsection{Occupancy test}
One of the crucial parameters in design of pixel readout chip is occupancy.
It is especially important for chips close to the beam line where track density
is higher. 

\begin{figure}[htb] 
\centerline{\epsfig{figure=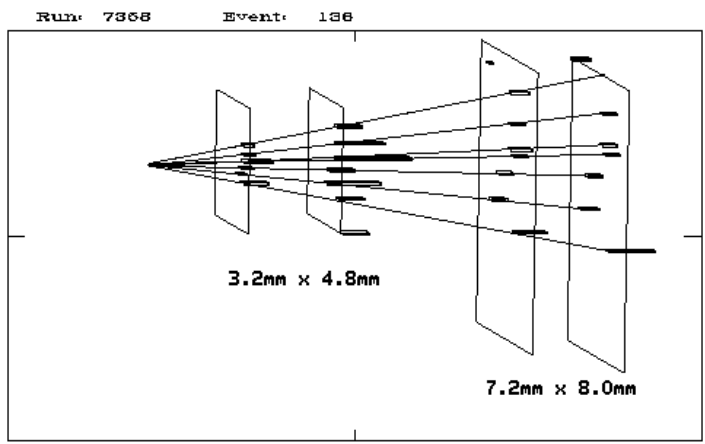,height=2.5in}}
\vspace{-.5cm}
\caption{ \label{diamond} Multi-particle event recorded from 4 plane pixel telescope, with
an upstream  diamond target.}
\end{figure}

In a short test, a diamond target was placed upstream of the four plane pixel
telescope. 
Multi-particle interactions were recorded and analyzed (see Figure~\ref{diamond}).
The track density is factor of 10 larger than expected at BTeV.

\section{Summary}
A test beam has shown that the pixel technology chosen for BTeV performs
according to expectation.
Two generations of pixel readout chips FPIX0 and FPIX1 are used, with
ATLAS sensor prototypes bump-bonded to them.
The spatial resolution with 2-bit analog information is quite good, and demonstrates
that the final version of the front end electronics featuring 3 bit FADC will provide
an excellent performance that will allow us to achieve our physics goals.

\end{document}